\documentclass[preprint,12pt]{elsarticle}  
\usepackage{graphicx} 
\usepackage{amssymb} 
\usepackage{amsmath} 

\journal{Physics Letters B} 

\def\lamb#1#2{$^{#1}_{\Lambda}${#2}}
\def\lam#1#2{$^{#1}_{~\Lambda}${#2}}
\def\la#1#2{$^{#1}_{~~\Lambda}${#2}} 

\begin{document} 

\begin{frontmatter} 

\title{Neutron-rich hypernuclei: $_{\Lambda}^{6}$H and beyond} 

\author[a]{A.~Gal\corref{cor1}} 
\cortext[cor1]{Corresponding author: Avraham Gal, avragal@vms.huji.ac.il} 
\author[b]{D.J.~Millener} 
\address[a]{Racah Institute of Physics, The Hebrew University, 91904 
Jerusalem, Israel} 
\address[b]{Physics Department, Brookhaven National Laboratory, Upton, 
NY 11973, USA} 

\date{\today} 

\begin{abstract} 
Recent experimental evidence presented by the FINUDA Collaboration for 
a particle-stable $_{\Lambda}^{6}$H has stirred renewed interest in charting 
domains of particle-stable neutron-rich $\Lambda$ hypernuclei, particularly 
for unbound nuclear cores. We have studied within a shell-model approach 
several neutron-rich $\Lambda$ hypernuclei in the nuclear $p$ shell that 
could be formed in $(\pi^-,K^+)$ or in $(K^-,\pi^+)$ reactions on stable 
nuclear targets. Hypernuclear shell-model input is taken from a theoretically 
inspired successful fit of $\gamma$-ray transitions in $p$-shell $\Lambda$ 
hypernuclei which includes also $\Lambda N \leftrightarrow \Sigma N$ coupling 
($\Lambda\Sigma$ coupling). The particle stability of $_{\Lambda}^{6}$H is 
discussed and predictions are made for binding energies of $_{\Lambda}^{9}$He, 
$_{~\Lambda}^{10}$Li, $_{~\Lambda}^{12}$Be, $_{~\Lambda}^{14}$B. None of the 
large effects conjectured by some authors to arise from  $\Lambda\Sigma$ 
coupling is borne out, neither by these realistic $p$-shell calculations, 
nor by quantitative estimates outlined for heavier hypernuclei with 
substantial neutron excess. 
\end{abstract} 


\end{frontmatter} 

\section{Introduction} 
\label{sec:intro} 

Dalitz and Levi Setti, fifty years ago \cite{DLS63}, discussed the possibility 
that $\Lambda$ hyperons could stabilize particle-unstable nuclear cores of 
$\Lambda$ hypernuclei and thus allow studies of neutron-rich baryonic systems 
beyond the nuclear drip line. The $\Lambda$'s effectiveness to enhance binding 
is primarily connected with the Pauli principle from which it is exempt, 
allowing it to occupy the lowest $0s_{\Lambda}$ orbital. Several unbound-core 
$\Lambda$ hypernuclei have since been identified in emulsion work, 
the neutron richest of which is $_{\Lambda}^{8}$He \cite{davis05}. 
Of particular interest is the recent FINUDA evidence for a particle-stable 
$_{\Lambda}^{6}$H hypernucleus produced in the $^6$Li($K^{-}_{\rm stop},
\pi^{+}$) reaction \cite{fingal12prl,fingal12npa}. In distinction from the 
well-established hyper-hydrogen isotopes $_{~\Lambda}^{3,4}$H, the $^5$H 
nuclear core of $_{\Lambda}^{6}$H is unbound, and its neutron-proton excess 
ratio $(N-Z)/(N+Z)$=0.6 is unsurpassed by any stable or $\Lambda$-stabilized 
core nucleus. The $_{\Lambda}^6$H hypernucleus was highlighted in 
Ref.~\cite{akaishi99} as a testground for the significance of $\Lambda\Sigma$ 
coupling in $\Lambda$ hypernuclei, spurred by the role it plays in $s$-shell 
hypernuclei~\cite{akaishi00,nemura02} and by the far-reaching consequences it 
might have for dense neutron-star matter with strangeness \cite{shinmura02}. 
In the present work, and as a prelude to its main theme, we discuss the 
particle stability of \lamb{6}{H} from the point of view of the shell model, 
focusing on $\Lambda\Sigma$ coupling contributions and making comparisons with 
other calculations and with the FINUDA experimental evidence. 

The purpose of this Letter is to provide shell-model predictions for other 
neutron-rich $\Lambda$ hypernuclei that could be reached at J-PARC in 
$(\pi^-,K^+)$ or $(K^-,\pi^+)$ reactions on stable nuclear targets in the $p$ 
shell. A missing-mass $^6$Li($\pi^{-},K^{+}$) spectrum from J-PARC in-flight 
experiment E10 \cite{E10} is under study at present, aimed at assessing 
independently FINUDA's evidence for a particle-stable $_{\Lambda}^{6}$H 
hypernucleus. In the present analysis, based on extensive hypernuclear 
shell-model calculations \cite{millener12}, we use $0p_N0s_{\Lambda}$ 
effective interactions with matrix elements constrained by the most 
comprehensive set of hypernuclear $\gamma$-ray measurements \cite{tamura10}. 
Included explicitly are also $0p_N0s_{\Lambda}\leftrightarrow 0p_N0s_{\Sigma}$ 
effective interactions based on the $0s_N0s_{Y}$ $G$-matrix interactions 
($Y$=$\Lambda,\Sigma$) used in the comprehensive $s$-shell hypernuclear 
calculations of Ref.~\cite{akaishi00}. The methodology of this shell-model 
analysis is briefly reviewed in Section~\ref{sec:meth}, following which 
we discuss $_{\Lambda}^{6}$H in Section~\ref{sec:L6H} and then, in 
Section~\ref{sec:nrich}, the neutron-rich hypernuclei that can be produced 
on stable nuclear targets in the nuclear $p$ shell. Predictions are made for 
the corresponding ground-state binding energies . Also outlined in this 
section is a shell-model evaluation of $\Lambda\Sigma$ coupling effects in 
heavier hypernuclei with larger neutron excess, such as \lam{49}{Ca} and 
\la{209}{Pb}, demonstrating that the increase in neutron excess is more 
than compensated by the decrease of the $\Lambda\Sigma$ coupling matrix 
elements with increasing orbital angular momentum $\ell_N$ of the valent 
nucleon configurations involved in the coherent coupling approximation. 
We thus conclude that none of the large effects conjectured in 
Ref.~\cite{akaishi99} to arise from $\Lambda\Sigma$ coupling is borne out 
in realistic calculations.

\section{Shell-model methodology} 
\label{sec:meth} 

The $\Lambda N$ effective interaction 
\begin{equation} 
V_{\Lambda N}={\bar V}+\Delta{\vec s}_N\cdot{\vec s}_{\Lambda}+
{\cal S}_{\Lambda}{\vec l}_N\cdot{\vec s}_{\Lambda}+{\cal S}_N{\vec l}_N\cdot
{\vec s}_N+{\cal T}S_{N\Lambda}, 
\label{eq:V_LN} 
\end{equation} 
where $S_{N\Lambda}=3({\vec\sigma}_N\cdot{\vec r})({\vec\sigma}_{\Lambda}
\cdot{\vec r})-{\vec\sigma}_N\cdot{\vec\sigma}_{\Lambda}$, is specified here 
by its $0p_N0s_{\Lambda}$ spin-dependent matrix elements within a $0\hbar
\omega$ shell-model space \cite{GSD71}. The same parametrization applies also 
to the $\Lambda\Sigma$ coupling interaction and the $\Sigma N$ interaction for 
both isospin 1/2 and 3/2, with an obvious generalization to account for the 
isospin 1 of the $\Sigma$ hyperon \cite{millener12}. The detailed properties 
of the $\Sigma N$ interaction parameters hardly matter in view of the large 
energy denominators of order $M_{\Sigma}-M_{\Lambda}\approx 80$ MeV with 
which they appear. To understand the effects of the $\Lambda\Sigma$ 
coupling interaction, it is convenient to introduce an overall isospin 
factor $\sqrt{4/3}\;{\vec t}_N\cdot{\vec t}_{\Lambda\Sigma}$, where 
${\vec t}_{\Lambda\Sigma}$ converts a $\Lambda$ to $\Sigma$ in isospace. 
Matrix elements of the $\Lambda N$ effective interaction (\ref{eq:V_LN}) 
and of the $\Lambda\Sigma$ coupling interaction are 
listed in Table~\ref{tab:V_YN}. The $\Lambda N$ matrix elements were fitted 
to a wealth of hypernuclear $\gamma$-ray measurements~\cite{millener10}, 
resulting in values close to those derived from the $YN$ interaction models 
NSC97~\cite{NSC97}. The $\Lambda\Sigma$ matrix elements 
are derived from the same NSC97 models used to construct $0s_N0s_{Y}$ 
$G$-matrix interactions in $s$-shell $\Lambda$ hypernuclear calculations 
\cite{akaishi00}. By limiting here the $\Sigma N$ model-space to 
$0p_N0s_{\Sigma}$, in parallel to the $0\hbar\omega$ $0p_N0s_{\Lambda}$ 
model-space used for $\Lambda N$, we maintain the spirit of Akaishi's 
{\it coherent} approximation \cite{akaishi99,akaishi00} which is designed 
to pick up the strongest $\Lambda\Sigma$ matrix elements.{\footnote{By 
{\it coherent} we mean $nl_N0s_{\Lambda}\leftrightarrow nl_N0s_{\Sigma}$ 
coupling that preserves hyperon and nucleon orbits, where $nl_N$ denotes 
the nuclear orbits occupied in the core-nucleus wavefunction. For the $A=4$ 
hypernuclei considered by Akaishi et al. \cite{akaishi00}, this definition 
reduces to the assumption that the $\Lambda$ and $\Sigma$, both in their 
$0s$ orbits, are coupled to the {\it same} nuclear core state. The parameters 
listed in the last line of Table~\ref{tab:V_YN} contribute almost 0.6 MeV to 
the binding of \lamb{4}{H}($0^+_{\rm g.s.}$) and more than 0.5 MeV to the 
excitation energy of \lamb{4}{H}($1^+_{\rm exc.}$).}} 

\begin{table}[thb] 
\begin{center} 
\caption{$0l_N 0s_{\Lambda} \leftrightarrow 0l_N 0s_Y$ matrix elements 
(in MeV) from Ref.~\cite{millener12}. $\bar{V}$ and $\Delta$ are the 
spin-average and difference of the triplet and singlet central matrix 
elements. For the $s_N^3s_Y$ hypernuclei, the $\Lambda\Sigma$ coupling matrix 
elements are $v(0^+_{\rm g.s.})={\bar V}^{0s}_{\Lambda\Sigma}+\frac{3}{4}
\Delta^{0s}_{\Lambda\Sigma}$ and $v(1^+_{\rm exc.})={\bar V}^{0s}_{\Lambda
\Sigma}-\frac{1}{4}\Delta^{0s}_{\Lambda\Sigma}$, with downward energy shifts 
$\delta E_{\downarrow}(J^{\pi})\approx v^2(J^{\pi})/(80~{\rm MeV})$.} 
\begin{tabular*}{\textwidth}{@{}l@{\extracolsep{\fill}}cccccc}
\hline\noalign{\smallskip}  
$Y$ & $0l_N$ & $\bar{V}$ & $\Delta$ & ${\cal S}_{\Lambda}$ & ${\cal S}_N$ & 
${\cal T}$ \\ 
\noalign{\smallskip}\hline\noalign{\smallskip}
$\Lambda$ ($A\leq 9$) & $0p_N$ & & 0.430 & $-$0.015 & $-$0.390 & 0.030 \\ 
$\Lambda$ ($A\geq 10$)& $0p_N$ & & 0.330 & $-$0.015 & $-$0.350 & 0.024 \\ 
$\Sigma$ & $0p_N$ & 1.45 & 3.04 &  $-$0.09 & $-$0.09 & 0.16 \\ 
$\Sigma$& $0s_N$ & 2.96 & 5.09 & -- & -- & -- \\ 
\noalign{\smallskip}\hline
\end{tabular*} 
\label{tab:V_YN} 
\end{center} 
\end{table} 

It is clear from Table~\ref{tab:V_YN} that the only significant 
$\Lambda\Sigma$ interaction parameters are ${\bar V}_{\Lambda\Sigma}$ and 
$\Delta_{\Lambda\Sigma}$, in obvious notation. The first one is associated 
with diagonal matrix elements of the spin-independent part of the 
$\Lambda\Sigma$ interaction, viz. 
\begin{equation} 
\langle (J_N T, s_{\Lambda})JT|V_{\Lambda\Sigma}|(J_N T, s_{\Sigma})JT \rangle 
= \sqrt{4/3}\;\sqrt{T(T+1)}\;{\bar V}_{\Lambda\Sigma}. 
\label{eq:V_LS} 
\end{equation} 
This part preserves the nuclear core, specified here by its total angular 
momentum $J_N$ and isospin $T$, with matrix elements that bear resemblance 
to the Fermi matrix elements in $\beta$ decay of the core nucleus. 
Similarly, the spin-spin part of the $\Lambda\Sigma$ interaction associated 
with the matrix element $\Delta_{\Lambda\Sigma}$ involves the operator 
$\sum_j{{\vec s}_{Nj}{\vec t}_{Nj}}$ for the core, connecting core states 
with large Gamow-Teller (GT) transition matrix elements as emphasized recently 
by Umeya and Harada in their study of the $_{~~~\Lambda}^{7-10}$Li 
isotopes~\cite{umeya11}. 

Finally, the $\Lambda N$ spin-independent matrix element $\bar{V}$ is not 
specified in Table~\ref{tab:V_YN} because it is not determined by fitting 
hypernuclear $\gamma$-ray transitions. Its value is to be deduced from fitting 
absolute binding energies. Suffice to say that $\bar{V}$ assumes values of 
$\bar{V}\approx -1.0\pm 0.1$~MeV \cite{millener12}. We consider it as part 
of a mean-field description of $\Lambda$ hypernuclei, consistent with the 
observation that on the average throughout the $p$ shell $B_{\Lambda}(A)$ 
increases by about 1 MeV upon increasing $A$ by one unit, with a 0.1 MeV 
uncertainty that might reflect genuine $YNN$ three-body contributions 
\cite{GSD78,MGDD85}.

\section{$_{\Lambda}^{6}$H} 
\label{sec:L6H} 

\begin{figure}[thb] 
\begin{center} 
\includegraphics[width=0.9\textwidth]{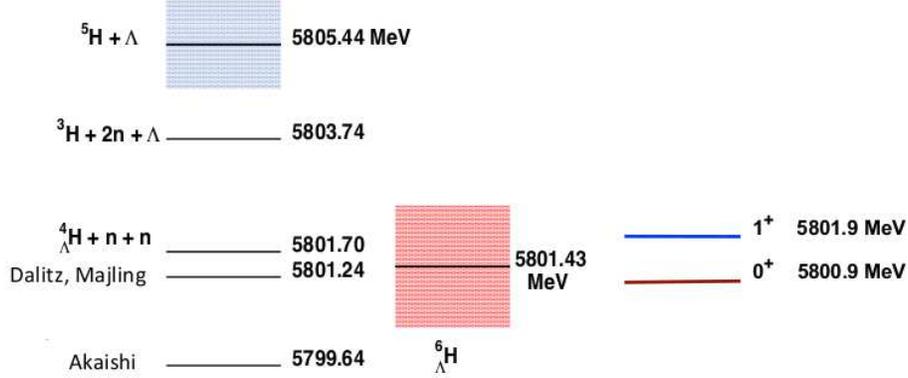} 
\caption{\lamb{6}{H} binding energy diagram adapted from 
Ref.~\cite{fingal12npa}, with thresholds marked on its upper-left side and 
theory predictions beneath \cite{DLS63,myint02}. The location of the $^{5}$H 
resonance vs. the $^{3}$H+$2n$ threshold, with the (blue) hatched box 
denoting its width, is taken from Ref.~\cite{korshen01}. The (red) shaded box 
represents the error on the mean production and decay mass value obtained from 
the three FINUDA events assigned to \lamb{6}{H}, whereas the positions of the 
$1^{+}_{\rm exc.}$ and $0^{+}_{\rm g.s.}$ levels marked on the right-hand side 
are derived separately from production and decay, respectively. We thank Elena 
Botta for providing this figure.} 
\label{fig:L6H} 
\end{center} 
\end{figure} 

The low-lying spectrum of \lamb{6}{H} consists of a $0^{+}_{\rm g.s.}$ and 
$1^{+}_{\rm exc.}$ states which, disregarding the two $p$-shell neutrons 
known from $^{6}$He to couple dominantly to $L\!=\!S\!=\!0$, are split in 
\lamb{4}{H} by 1.08$\pm$0.02 MeV \cite{tamura12}. These states in \lamb{6}{H} 
are split by 1.0$\pm$0.7 MeV, judging by the systematic difference noted in 
the FINUDA experiment \cite{fingal12prl,fingal12npa} between the mass values 
$M$(\lamb{6}{H}) from production and decay, as shown in Fig.~\ref{fig:L6H}. 
The observation of \lamb{6}{H}$\to\pi^{-}+{^6}$He weak decay in the FINUDA 
experiment implies that \lamb{6}{H} is particle-stable, with $2n$ separation 
energy $B_{2n}({_{\Lambda}^{6}\rm{H_{g.s.}}})$=0.8$\pm$1.2 MeV, independently 
of the location of the core-nucleus $^5$H resonance. 

\subsection{Phenomenological shell-model analysis} 
\label{subsec:phenomen} 

A phenomenological shell-model estimate for $B_{\Lambda}({_{\Lambda}^{6}{
\rm H}})$ cited in Ref.~\cite{fingal12npa}, also bypassing $^5$H, yields 
\begin{equation} 
B_{\Lambda}({_{\Lambda}^{6}{\rm H}}) = B_{\Lambda}({_{\Lambda}^{4}{\rm H}}) + 
[B_{\Lambda}({_{\Lambda}^{7}{\rm He}})-B_{\Lambda}({_{\Lambda}^{5}{\rm He}})] 
= (2.04\pm 0.04) + (2.24\pm 0.09)~{\rm MeV}, 
\label{eq:BL} 
\end{equation} 
based on the value $B_{\Lambda}({_{\Lambda}^{7}{\rm He}})=(5.36\pm 0.09)$ 
MeV obtained by extrapolating linearly the known binding energies of the 
other $T=1$ isotriplet members of the $A=7$ hypernuclei.{\footnote{The value 
$B_{\Lambda}({_{\Lambda}^{7}{\rm He}})$=5.68$\pm$0.03(stat.)$\pm$0.22(syst.) 
MeV, obtained recently from Jlab E01-011 \cite{nue13}, is irreproducible 
also in related few-body cluster calculations \cite{hiyama09}.}} The argument 
underlying this derivation is that the total $\Lambda nn$ interaction, 
including $\Lambda\Sigma$ coupling contributions, is given by the difference 
of $\Lambda$ binding energies within the square bracket, assuming that the 
$\Lambda nn$ configuration for the two $p$-shell neutrons in \lamb{6}{H} 
is identical with that in \lamb{7}{He}. Accepting then the $^{5}$H mass 
determination from Ref.~\cite{korshen01} one obtains the corresponding 
phenomenological shell-model estimate for $B_{2n}$(\lamb{6}{H}): 
\begin{equation} 
B_{2n}({_{\Lambda}^{6}{\rm H}}) = B_{2n}({^{5}{\rm H}}) + 
[B_{\Lambda}({_{\Lambda}^{6}{\rm H}})-B_{\Lambda}({_{\Lambda}^{4}{\rm H}})] = 
(-1.7\pm 0.3) + (2.24\pm 0.09)~{\rm MeV},  
\label{eq:B2n} 
\end{equation} 
yielding $B_{2n}({_{\Lambda}^{6}{\rm H}})$=(0.5$\pm$0.3) MeV which, however, 
is not independent of the resonance location of $^5$H. Equation~(\ref{eq:B2n}) 
demonstrates the gain in {\it nuclear} binding owing to the added $\Lambda$ 
hyperon. 
\subsection{Refined shell-model analysis} 
\label{subsec:refined} 

The phenomenological shell-model estimate outlined above needs to be refined 
on two counts as follows. First, one notes that the nuclear $s$ shell is not 
closed in \lamb{6}{H}, in distinction from all other neutron-rich $\Lambda$ 
hypernuclei considered in the present work, so that the $\Lambda\Sigma$ 
coupling contribution in \lamb{6}{H} is not the sum of the separate 
contributions from the $s$ shell through \lamb{4}{H} and from the $p$ shell 
through \lamb{7}{He}. For an $s^3p^2$ core of \lamb{6}{H}, with [32] spatial 
symmetry and $T\!=\!3/2$, the coherent $\Lambda\Sigma$ coupling diagonal 
matrix element is $\sqrt{5/9}\,v(J^{\pi})+\sqrt{20/9}\,\bar{V}^{0p}_{\Lambda
\Sigma}$ for $J^{\pi}\!=\!0^+,\!1^+$, where $v(J^{\pi})$ is the corresponding 
$A\!=\!4$ matrix element (see caption to Table~\ref{tab:V_YN}), with no 
contribution from $\Delta^{0p}_{\Lambda\Sigma}$ because the two $p$-shell 
neutrons have Pauli spin zero. $\Lambda\Sigma$ contributions to the binding 
energies of $\Lambda$ hypernuclei involved in the shell-model evaluation of 
the \lamb{6}{H} doublet levels $0^+_{\rm g.s.}$ and $1^+_{\rm exc.}$, 
calculated for the interaction specified in Table~\ref{tab:V_YN}, are listed 
in Table~\ref{tab:LamSig}. 

\begin{table}[hbt] 
\begin{center} 
\caption{$\Lambda\Sigma$ contributions (in keV) to binding energies of 
$\Lambda$ hypernuclei involved in shell-model considerations of \lamb{6}{H} 
stability.} 
\begin{tabular*}{\textwidth}{@{}l@{\extracolsep{\fill}}ccccc} 
\hline\noalign{\smallskip} 
$J^{\pi}$& \lamb{4}{H} & \lamb{7}{He}(${\frac{1}{2}}^+_{\rm g.s.}$) 
& \lamb{7}{He}(${\frac{1}{2}}^+_{\rm g.s.}$) & \lamb{6}{H} & \lamb{6}{H} \\ 
         & diag.=tot. & diag. & total & diag. & total \\
\noalign{\smallskip}\hline\noalign{\smallskip} 
0$^+$ & 574 & 70 & 98 & 650 & 821 \\ 
1$^+$ &  36 & 70 & 98 & 146 & 176 \\ 
\noalign{\smallskip}\hline 
\end{tabular*} 
\label{tab:LamSig} 
\end{center} 
\end{table} 

Whereas \lamb{4}{H} admits only a diagonal matrix-element contribution, 
the $p$-shell hypernuclei admit both diagonal as well as nondiagonal 
matrix-element contributions, with the listed diagonal ones dominating. 
It is also seen that \lamb{6}{H}($0^+_{\rm g.s.}$) gains 149 keV and 
\lamb{6}{H}($1^+_{\rm exc.}$) gains 42 keV binding above the sum of 
separate contributions from \lamb{4}{H} and \lamb{7}{He} inherent in the 
phenomenological shell-model approach. With respect to \lamb{4}{H} alone, the 
$\Lambda\Sigma$ contribution gain is 247 keV for \lamb{6}{H}($0^+_{\rm g.s.}$) 
and 140 keV for \lamb{6}{H}($1^+_{\rm exc.}$). Thus, the shell-model predicts 
a doublet spacing $\Delta E$(1$^{+}_{\rm exc.}$--0$^{+}_{\rm g.s.}$) in 
\lamb{6}{H} of only 0.1 MeV larger than in \lamb{4}{H} owing to the 
$\Lambda\Sigma$ coupling, considerably less than the additional 1.4 MeV 
argued by Akaishi et al. \cite{akaishi99,myint02}. 

A more important reason for revising the phenomenological shell-model estimate 
is the reduction of matrix elements that involve spatially extended $p$-shell 
neutrons in \lamb{6}{H} ($B_{2n}\lesssim 0.8$ MeV) relative to matrix elements 
that involve a more compact $p$-shell neutron in \lamb{7}{He} ($B_n\sim 3.0$ 
MeV). The average neutron separation energy in \lamb{6}{H} is closer to that 
in \lamb{6}{He} ($B_n = 0.26\pm 0.10$ MeV). Using our shell-model estimates 
for the spin-dependent and $\Lambda\Sigma$ coupling contributions to 
$B_{\Lambda}$(\lamb{6}{He}), we deduce $\bar{V}\sim -0.8$~MeV and a similar 
$20\%$ reduction in other $0p_N0s_{\Lambda}$ interaction matrix elements, 
thereby yielding 
\begin{equation} 
B_{\Lambda}({_{\Lambda}^{6}{\rm H}}) \sim B_{\Lambda}({_{\Lambda}^{4}{\rm H}}) 
+ 0.8[B_{\Lambda}({_{\Lambda}^{7}{\rm He}})-B_{\Lambda}({_{\Lambda}^{5}{\rm 
He}})] = 
(3.83\pm 0.08\pm 0.22)~{\rm MeV}, 
\label{eq:B_Lphen} 
\end{equation} 
where the first uncertainty is due to the statistical uncertainty of the 
binding-energy input \cite{davis05} and the second, systematic uncertainty 
assumes that the 0.8 renormalization factor is uncertain to within 0.1. 
Accepting, again, the $^5$H resonance location from \cite{korshen01}, 
a revised estimate follows: $B_{2n}$(\lamb{6}{H})$\approx$(0.1$\pm$0.4) MeV. 
Because the $s^3p^2$ core for \lamb{6}{H} will be more spatially extended 
than the $\alpha$-particle core for the He hypernuclei, a few-body calculation 
such as that of Hiyama et al.~\cite{hiyama13}, discussed below, is 
required.{\footnote{Comparing $p$-shell neutron densities for a slightly 
bound \lamb{6}{H} \cite{hiyama13} with those for \lamb{7}{He} \cite{hiyama09} 
in four-body C-$n$-$n$-$\Lambda$ calculations, where the cluster C stands 
for $^3$H or $^4$He respectively, supports our $\approx$20\% reduction of 
$\Lambda N$ and $\Lambda\Sigma$ matrix elements in going from \lamb{7}{He} 
to \lamb{6}{H}. We thank Emiko Hiyama for providing us with plots of these 
densities.}} We therefore consider the present revised shell-model estimate, 
listed also in Table~\ref{tab:L6H} below, as an upper bound for 
$B_{\Lambda}$(\lamb{6}{H}). 

\begin{table}[thb] 
\begin{center} 
\caption{\lamb{6}{H}($0^{+}_{\rm g.s.}$) binding energies $B_{2n/\Lambda}$ 
and doublet spacing $\Delta E$(1$^{+}_{\rm exc.}$--0$^{+}_{\rm g.s.}$) 
predictions vs. experiment. $E_{2n}$ marks the assumed $^5$H resonance position 
w.r.t. $2n$ emission.} 
\begin{tabular*}{\textwidth}{@{}l@{\extracolsep{\fill}}ccccc} 
\hline 
$B$ \& $\Delta E$ & \multicolumn{2}{c}{Akaishi} & shell model & Hiyama & 
FINUDA \\ 
(MeV) & \cite{akaishi99} & \cite{myint02} & this work & \cite{hiyama13} & 
exp.~\cite{fingal12prl,fingal12npa} \\ 
\hline\noalign{\smallskip} 
$E_{2n}$($^5$H) & 2.1 & 1.7 & 1.7$\pm$0.3 & ~~1.6 & -- \\ 
$B_{2n}$(\lamb{6}{H}) & 1.7 & 2.1 & 0.1$\pm$0.4 & $-$1.1 & 0.8$\pm$1.2 \\ 
$B_{\Lambda}$(\lamb{6}{H}) & 5.8 & 5.8 & 3.8$\pm$0.2 & ~~2.5 & 4.5$\pm$1.2 \\ 
$\Delta E$(\lamb{6}{H}) & 2.4 & 2.4 & 1.2$\pm$0.1 &--& 1.0$\pm$0.7 \\ 
\noalign{\smallskip}\hline 
\end{tabular*} 
\label{tab:L6H} 
\end{center} 
\end{table} 

In Table~\ref{tab:L6H} we compare several theoretical \lamb{6}{H} predictions 
to each other and to FINUDA's findings. The table makes it clear that FINUDA's 
results do not support the predictions made in Refs.~\cite{akaishi99,myint02} 
according to which the two $p$-shell neutrons in \lamb{6}{H} enhance the 
$\Lambda\Sigma$ coupling contribution in \lamb{4}{H}, thereby pushing down 
the \lamb{6}{H}($0^{+}_{\rm g.s.}$) level by twice as much as it does in 
\lamb{4}{H} and thus doubling the $0_{\rm g.s.}^+$--$1_{\rm exc.}^+$ spacing. 
In contrast to such overbound \lamb{6}{H}, the recent $^{3}$H-$n$-$n$-$
\Lambda$ four-body calculation by Hiyama et al. \cite{hiyama13} does not bind 
\lamb{6}{H}. This calculation is the only one that considers dynamically the 
$^5$H core as a three-body $^{3}$H-$n$-$n$ resonance, but then it disregards 
$\Lambda\Sigma$ coupling which is a necessary ingredient in any \lamb{6}{H} 
calculation owing to the lowest particle-stability threshold which 
involves \lamb{4}{H}. Our own shell-model estimate, allowing approximately 
for a spatially extended $2n$ cluster, suggests a very weakly bound 
\lamb{6}{H}($0^{+}_{\rm g.s.}$) and a particle-unstable 
\lamb{6}{H}($1_{\rm exc.}^+$) that decays by emitting a low-energy neutron 
pair:  
\begin{equation} 
{_{\Lambda}^{6}{\rm H}}(1_{\rm exc.}^+) \rightarrow 
{_{\Lambda}^{4}{\rm H}}(0^{+}_{\rm g.s.})+2n. 
\label{eq:decay} 
\end{equation}  
However, this decay is substantially suppressed both kinematically and 
dynamically, kinematically since $s$-wave emission requires a $^3S_1$ 
dineutron configuration which is Pauli-forbidden, and the allowed $p$-wave 
emission which is kinematically suppressed at low energy requires that 
{\it both} \lamb{6}{H} constituents, \lamb{4}{H} and $2n$, flip their 
Pauli spin which is disfavored dynamically. This leaves open the possibility 
that $M1$ $\gamma$ emission to \lamb{6}{H}($0^{+}_{\rm g.s.}$) provides 
a competitive decay mode of the $1_{\rm exc.}^+$ level.

\section{Neutron-rich hypernuclei beyond \lamb{6}{H}} 
\label{sec:nrich} 

In the first part of this section we consider neutron-rich $\Lambda$ 
hypernuclei that can be produced in double-charge exchange reactions 
$(\pi^-,K^+)$ or $(K^-,\pi^+)$ on stable nuclear targets in the $p$ shell. 
These are \lamb{6}{H}, which was discussed in the previous section, 
\lamb{9}{He}, \lam{10}{Li}, \lam{12}{Be}, and \lam{14}{B}, with targets 
$^6$Li, $^9$Be, $^{10}$B, $^{12}$C, and $^{14}$N, respectively.{\footnote{We 
excluded \lam{16}{C} which can be produced on $^{16}$O target because its 
analysis involves $(1s-0d)_N 0s_{\Lambda}$ matrix elements which are not 
constrained by any hypernuclear datum. For the same reason we ignored for 
\lam{12}{Be} the positive-parity doublet built on $^{11}$Be$_{\rm g.s}({
\frac{1}{2}}^+)$, considering only the normal-parity doublet based on the 
${\frac{1}{2}}^-$ excited state at 0.32 MeV.}} In distinction from the 
unbound-core \lamb{6}{H}, all other neutron-rich $\Lambda$ hypernuclei 
listed above are based on bound nuclear cores, which ensures their particle 
stability together with that of a few excited states. In the second half of 
the present section we outline the evaluation of $\Lambda\Sigma$ coupling 
matrix elements in heavier nuclear cores with substantial neutron excess, 
$^{48}$Ca and $^{208}$Pb.

\subsection{$p$-shell neutron-rich hypernuclei beyond \lamb{6}{H}} 
\label{subsec:pshell} 

\begin{table}[thb]  
\begin{center} 
\caption{Beyond-mean-field $\Delta B^{\rm g.s.}_{\Lambda}$ shell-model 
contributions to normal-parity ground states of neutron-rich hypernuclei 
(in MeV); see text. ME stands for ``matrix element".} 
\begin{tabular*}{\textwidth}{@{}c@{\extracolsep{\fill}}cccccc} 
\hline\noalign{\smallskip} 
target & $n$--rich & ME & $\Lambda\Sigma$ & $\Lambda\Sigma$ & 
induced & $\Delta B^{\rm g.s.}_{\Lambda}$ \\ 
$^{A}Z$ &  $_{\Lambda}^{A}(Z-2)$ & diag. & diag. & total & 
${\vec l}_N\cdot {\vec s}_N$ & total \\ 
\noalign{\smallskip}\hline\noalign{\smallskip} 
$^{9}$Be & \lamb{9}{He}(${\frac{1}{2}}^+$) & 4.101 & 0.210 & 0.253 & 
0.619 & 0.879 \\ 
$^{10}$B & \lam{10}{Li}($1^-$) & 4.023 & 0.202 & 0.275 & 0.595 & 1.022 \\ 
$^{12}$C & \lam{12}{Be}($0^-$) & 3.835 & 0.184 & 0.158 & 0.554 & 0.748 \\ 
$^{14}$N & \lam{14}{B}($1^-$)  & 3.884 & 0.189 & 0.255 & 0.458 & 0.785 \\ 
\noalign{\smallskip}\hline 
\end{tabular*} 
\label{tab:pshellres} 
\end{center}  
\end{table} 

In Table~\ref{tab:pshellres}, we focus on $\Lambda\Sigma$ contributions to 
binding energies of normal-parity g.s. in neutron-rich $p$-shell hypernuclei. 
Column 3 gives the diagonal matrix-element (ME) from the central components 
${\bar V}^{0p}_{\Lambda\Sigma}$ and $\Delta^{0p}_{\Lambda\Sigma}$ of the 
coherent $\Lambda\Sigma$ coupling interaction. The contribution from 
${\bar V}^{0p}_{\Lambda\Sigma}$ is given by Eq.~(\ref{eq:V_LS}), saturating 
this diagonal matrix element in \lamb{9}{He} and dominating it with a value 
3.242 MeV in the other cases. Column 4 gives the corresponding downward 
energy shift (ME)$^2$/80 assuming a constant 80 MeV separation between 
$\Lambda$ and $\Sigma$ states, and column 5 gives the energy shift in the 
full shell-model calculation, accounting also for the noncentral components 
of the coherent $\Lambda\Sigma$ coupling interaction and including nondiagonal 
matrix elements as well. Except for \lam{12}{Be} where the decreased shift 
in the complete calculation is due to the contribution of the noncentral 
components to the diagonal matrix element, a very good approximation for the 
total coherent $\Lambda\Sigma$ coupling contribution is obtained using just 
the central (Fermi and GT) $\Lambda\Sigma$ coupling interaction. In these 
cases, the increased energy shift beyond the diagonal contribution arises from 
$\Sigma$ core states connected to the $\Lambda$ core state by GT transitions. 
The shift gets smaller with increased fragmentation of the GT strength. 
Finally, we note that the total $\Lambda\Sigma$ contribution listed 
in column 5 agrees for \lam{10}{Li} with that computed in Ref.~\cite{umeya09} 
using the same $YN$ shell-model interactions. More recently, these authors 
discussed $\Lambda\Sigma$ coupling effects on g.s. doublet spacings in 
$_{~~~\Lambda}^{7-10}$Li isotopes, obtaining moderate enhancements between 
70 to 150 keV~\cite{umeya11}. We have reached similar conclusions for all the 
neutron-rich hypernuclei considered in the present work beyond \lamb{6}{H}. 

The total $\Lambda\Sigma$ contribution discussed above is only one component 
of the total beyond-mean-field (BMF) contribution $\Delta B^{\rm g.s.}_{
\Lambda}$. To obtain the latter, various spin-dependent $\Lambda N$ 
contributions generated by $V_{\Lambda N}$ have to be added to the total 
$\Lambda\Sigma$ contributions listed in column 5 of Table~\ref{tab:pshellres}. 
Column 6 lists one such spin-dependent $\Lambda N$ contribution, arising 
from the $\Lambda$-induced ${\vec l}_N\cdot{\vec s}_N$ nuclear spin-orbit 
term of Eq.~(\ref{eq:V_LN}). By comparing column 5 with column 6, and both 
with the total BMF contributions listed in column 7, the last one, it is 
concluded that the majority of the BMF contributions arise from $\Lambda N$ 
spin-dependent terms, dominated by ${\vec l}_N\cdot {\vec s}_N$. 

\begin{table}[thb] 
\begin{center} 
\caption{Binding energy predictions for neutron-rich $p$-shell hypernuclei 
(in MeV).} 
\begin{tabular*}{\textwidth}{@{}c@{\extracolsep{\fill}}cccccc} 
\hline\noalign{\smallskip} 
$n$-rich & normal & normal & normal & $n$-rich & $n$-rich \\ 
$_{\Lambda}^{A}Z$ & $_{\Lambda}^{A}Z'$ & $B^{\rm g.s.}_{\Lambda}$ & 
$\Delta B^{\rm g.s.}_{\Lambda}$ & $\Delta B^{\rm g.s.}_{\Lambda}$ & 
$B^{\rm g.s.}_{\Lambda}$ \\ 
\noalign{\smallskip}\hline\noalign{\smallskip} 
\lamb{9}{He}(${\frac{1}{2}}^+$) & \lamb{9}{Li}/\lamb{9}{B} & 8.44$\pm$0.10 & 
0.952 & 0.879 & 8.37$\pm$0.10 \\ 
\lam{10}{Li}($1^-$) & \lam{10}{Be}/\lam{10}{B} & 8.94$\pm$0.11 & 0.518 & 
1.022 & 9.44$\pm$0.11 \\ 
\lam{12}{Be}($0^-$) & \lam{12}{B} & 11.37$\pm$0.06 & 0.869 & 0.748 & 
11.25$\pm$0.06 \\ 
\lam{14}{B}($1^-$) & \lam{14}{C} & 12.17$\pm$0.33 & 0.904 & 0.785 & 
12.05$\pm$0.33 \\ 
\noalign{\smallskip}\hline 
\end{tabular*} 
\label{tab:B_L} 
\end{center} 
\end{table} 

Our $\Lambda$ binding energy predictions for ground states of neutron-rich 
hypernuclei $_{\Lambda}^{A}Z$ in the $p$ shell are summarized in the last 
column of Table~\ref{tab:B_L}. We used the experimentally known g.s. binding 
energies of same-$A$ normal hypernuclei $_{\Lambda}^{A}Z'$ with $Z'>Z$ 
from~\cite{davis05}, averaging statistically when necessary. 
The BMF contributions $\Delta B^{\rm g.s.}_{\Lambda}$(normal) to the 
binding energies of the $_{\Lambda}^{A}Z'$ `normal' hypernuclei, taken 
from~\cite{millener12}, are listed in column 4 and the BMF contributions 
$\Delta B^{\rm g.s.}_{\Lambda}$($n$-rich) to the binding energies of the 
$_{\Lambda}^{A}Z$ neutron-rich hypernuclei value, which are reproduced in 
the last column of Table~\ref{tab:pshellres}, are listed here in column 5. 
Finally, the predicted binding energies $B^{\rm g.s.}_{\Lambda}$($n$-rich) 
of the neutron-rich hypernuclei $_{\Lambda}^{A}Z$ are listed in the last 
column of Table~\ref{tab:B_L}, were evaluated according to 
\begin{equation} 
B^{\rm g.s.}_{\Lambda}({_{\Lambda}^{A}Z})=B^{\rm g.s.}_{\Lambda}
({_{\Lambda}^{A}Z'})-\Delta B^{\rm g.s.}_{\Lambda}({_{\Lambda}^{A}Z'})+
\Delta B^{\rm g.s.}_{\Lambda}({_{\Lambda}^{A}Z}). 
\label{eq:B_L} 
\end{equation} 
The resulting binding energies are lower by about 0.1 MeV than those of the 
corresponding normal hypernuclei, except for \lam{10}{Li} with binding energy 
about 0.5 MeV {\it larger} than that of \lam{10}{Be}--\lam{10}{B}. 

The relatively small effects induced by the $\Lambda\Sigma$ coupling 
interaction on $B^{\rm g.s.}_{\Lambda}$ persist also in the particle-stable 
portion of neutron-rich $\Lambda$ hypernuclear spectra. Here we only 
mention that, except for \lam{12}{Be}, one anticipates a clear separation 
about or exceeding 3 MeV between g.s. (in $_{\Lambda}^{9}$He) or g.s.-doublet 
(in $_{~\Lambda}^{10}$Li and $_{~\Lambda}^{14}$B) and the first-excited 
hypernuclear doublet. Thus, the g.s. or its doublet are likely to be observed 
in experimental searches of neutron-rich $\Lambda$ hypernuclei using $^9$Be, 
$^{10}$B, and $^{14}$N targets, provided a resolution of better than 2 MeV can 
be reached.

\subsection{Medium weight and heavy hypernuclei} 
\label{subsec:heavy} 

It was demonstrated in the previous subsection that the $\Lambda\Sigma$ 
coupling contributions to $p$-shell hypernuclear binding energies are 
considerably smaller than for the $s$-shell $A=4$ hypernuclei. The underlying 
$\Lambda\Sigma$ matrix elements decrease by roughly factor of two upon going 
from the $s$ shell to the $p$ shell, and the resulting energy contributions 
roughly by factor of four. This trend persists upon going to heavier 
hypernuclei and, as shown below, it more than compensates for the larger 
$N-Z$ neutron excess available in heavier hypernuclei. To demonstrate how 
it works, we outline schematic shell-model weak-coupling calculations using 
$G$-matrix $\Lambda\Sigma$ central interactions generated from the NSC97f $YN$ 
potential model by Halderson \cite{Hal}, with input and results listed in 
Table~\ref{tab:heavy} across the periodic table. We note that although the 
separate contributions $\Lambda\Sigma$($\bar V$) and $\Lambda\Sigma$($\Delta$) 
in \lamb{9}{He} differ from those using the Akaishi interaction 
(see Table~\ref{tab:pshellres}), the summed $\Lambda\Sigma$ contribution 
is the same to within $2\%$. For \lam{49}{Ca}, its $^{48}$Ca core consists 
of $0s,0p,1s-0d$ closed shells of protons and neutrons, plus a closed 
$0f_{7/2}$ neutron shell with isospin $T=4$. The required $0f_N 0s_{\Lambda}
\leftrightarrow 0f_N 0s_{\Sigma}$ radial integrals are computed for HO 
$n\ell_N$ radial wavefunctions with $\hbar\omega=45A^{-1/3}-25A^{-2/3}$ MeV 
and are compared in the table to the $0p_N 0s_{\Lambda} \leftrightarrow 
0p_N 0s_{\Sigma}$ radial integrals for \lamb{9}{He}. The decrease in the 
values of ${\bar V}_{\Lambda\Sigma}$ and of $\Delta_{\Lambda\Sigma}$ upon 
going from \lamb{9}{He} to \lam{49}{Ca} is remarkable, reflecting the 
poorer overlap between the hyperon $0s_Y$ and the nucleon $0\ell_N$ 
radial wavefunctions as $\ell_N$ increases. The resulting $\Lambda\Sigma$ 
contributions to the binding energy are given separately for the Fermi 
spin-independent interaction using the listed values of ${\bar V}_{\Lambda
\Sigma}$, and for the GT spin-dependent interaction using the listed values 
of $\Delta_{\Lambda\Sigma}$. The Fermi contribution involves a $0^+\to 0^+$ 
core transition preserving the value of $T$, see Eq.~(\ref{eq:V_LS}). 
The GT contribution consists of three separate $0^+\to 1^+$ core transitions 
of comparable strengths, transforming a $f_{7/2}$ neutron to (i) $f_{7/2}$ 
nucleon with nuclear-core isospin $T_N=3$, or to $f_{5/2}$ nucleon with (ii) 
$T_N=3$ or (iii) $T_N=4$. The overall $\Lambda\Sigma$ contribution of 24 keV 
in \lam{49}{Ca} is rather weak, one order of magnitude smaller than in 
\lamb{9}{He}. 

\begin{table}[htb] 
\begin{center} 
\caption{$\Lambda\Sigma$ matrix elements and contributions to binding 
energies (in MeV) for neutron-rich hypernuclei across the periodic table, 
using NSC97f $YN$ interactions from Halderson \cite{Hal}. In all cases,
the diagonal matrix element is given by $\sqrt{(T+1)/3T}\,\sum (2j+1)\,
\bar{V}_{\Lambda\Sigma}^{n\ell_N}$, where the sum runs over orbits in the 
neutron excess.} 
\begin{tabular*}{\textwidth}{@{}c@{\extracolsep{\fill}}cccccc} 
\hline\noalign{\smallskip} 
$N$$-$$Z$ & $_{\Lambda}^{A}Z$ & ${\bar V}_{\Lambda\Sigma}$ & 
$\Lambda\Sigma$($\bar V$) & $\Delta_{\Lambda\Sigma}$ & 
$\Lambda\Sigma$($\Delta$) & $\Delta B^{\rm g.s.}_{\Lambda}(\Lambda\Sigma)$ \\ 
\noalign{\smallskip}\hline\noalign{\smallskip} 
 4 & \lamb{9}{He} & 1.194  & 0.143 & 4.070 & 0.104 & 0.246 \\ 
 8 & \lam{49}{Ca} & 0.175  & 0.010 & 0.946 & 0.014 & 0.024 \\  
22 & \la{209}{Pb} & 0.0788 & 0.052 & 0.132 & 0.001 & 0.053 \\ 
\noalign{\smallskip}\hline 
\end{tabular*} 
\label{tab:heavy} 
\end{center} 
\end{table} 

In \la{209}{Pb}, the 44 excess neutrons run over the $2p$, $1f$, $0h_{9/2}$ 
and $0i_{13/2}$ orbits. The Fermi $0^+\to 0^+$ core transition is calculated 
using Eq.~(\ref{eq:V_LS}) where, again, ${\bar V}_{\Lambda\Sigma}$ 
(with a value listed in Table~\ref{tab:heavy}) stands for the $(2j+1)$ average 
of the separate neutron-excess ${\bar V}_{\Lambda\Sigma}^{n\ell_N}$ radial 
integrals, resulting in a $\Lambda\Sigma$ contribution of merely 52 keV to the 
binding energy of \la{209}{Pb}. For the calculation of the GT $0^+\to 1^+$ 
core transitions, we define $\Delta_{\Lambda\Sigma}$ as a $(2j+1)$ average of 
$\Delta_{\Lambda\Sigma}^{n\ell_N}$ radial integrals over the neutron-excess 
incomplete LS shells $0h_{9/2}$ and $0i_{13/2}$ (with a value listed in the 
table). These are the neutron orbits that initiate the GT core transitions 
required in \la{209}{Pb}, with structure similar to that encountered in \lam{
49}{Ca} for the $0f_{7/2}$ neutron orbit; details will be given elsewhere, 
suffice to say here that neither of these transitions results in a 
$\Lambda\Sigma$ contribution larger than 0.2 keV to the binding energy of 
\la{209}{Pb}.

\section{Summary and outlook} 
\label{sec:sum} 

In this Letter we have presented detailed shell-model calculations of 
$p$-shell neutron-rich $\Lambda$ hypernuclei using (i) $\Lambda N$ effective 
interactions derived by fitting to comprehensive $\gamma$-ray hypernuclear 
data, and (ii) theoretically motivated $\Lambda\Sigma$ 
interaction terms. None of the large effects conjectured by Akaishi and 
Yamazaki~\cite{akaishi99} to arise from $\Lambda\Sigma$ 
coherent coupling in neutron-rich hypernuclei is borne out by these realistic 
shell-model calculations. This is evident from the relatively modest 
$\Lambda\Sigma$ component of the total BMF contribution to 
the $\Lambda$ hypernuclear g.s. binding-energy, marked $\Delta B^{\rm g.s.}_{
\Lambda}$ in Table~\ref{tab:pshellres}. It should be emphasized, however, 
that $\Lambda\Sigma$ coupling plays an important role in doublet 
spacings in $p$-shell hypernuclei~\cite{millener12}, just as it does
for \lamb{4}{H} and \lamb{4}{He}~\cite{akaishi00}. Although the 
ground-state doublet spacings in \lam{10}{Li} and \lam{14}{B} probably
can't be measured, $\Lambda\Sigma$ coupling contributes 40\% and 55\%
to the predicted doublet spacings of 341 and 173 keV, respectively
(similar relative strengths as for \lam{16}{O}~\cite{millener12}; 
the \lam{12}{Be} doublet spacing, however, is predicted to be very small). 
Forthcoming experiments searching for neutron-rich $\Lambda$ hypernuclei 
at J-PARC will shed more light on the $N-Z$ dependence of hypernuclear 
binding energies. 

We have also discussed in detail the shell-model argumentation for a slightly 
particle-stable \lamb{6}{H}, comparing it with predictions by Akaishi et al. 
that overbind \lamb{6}{H} as well as with a very recent four-body calculation 
by Hiyama et al. that finds it particle-unstable. 
In the former case \cite{akaishi99,myint02}, we have argued that the effects 
of $\Lambda\Sigma$ coupling in \lamb{6}{H} cannot be very much larger than 
they are in \lamb{4}{H}, whereas such effects are missing in the latter 
case \cite{hiyama13}. Genuine $YNN$ three-body contributions are missing so 
far in {\it all} calculations of \lamb{6}{H}, and need to be implemented. 
Systematic studies of these contributions, which appear first at NNLO EFT 
expansions, have not been made to date in hypernuclear physics where the state 
of the art is just moving from LO to NLO applications \cite{haidenbauer13}. 
At present, and given the regularity of hypernuclear binding energies of 
heavy $\Lambda$ hypernuclei with respect to those in light and medium-weight 
hypernuclei as manifest in SHF-motivated density-dependent calculations 
\cite{MDG88}, in RMF \cite{MJ94} and in-medium chiral SU(3) dynamics 
calculations \cite{finelli09} extending over the whole periodic table, 
no strong phenomenological motivation exists to argue for substantial 
$N-Z$ effects in $\Lambda$ hypernuclei. Indeed, this was demonstrated 
in the calculation outlined in the previous section for \lam{49}{Ca} 
and \la{209}{Pb}.

\section*{Acknowledgements} 

We thank Emiko Hiyama and Ji\v{r}\'{i} Mare\v{s} for useful comments made 
on a previous version of this work. D.J.M. acknowledges the support by the 
U.S. DOE under Contract DE-AC02-98CH10886 with the Brookhaven National 
Laboratory, and A.G. acknowledges support by the EU initiative FP7, 
HadronPhysics3, under the SPHERE and LEANNIS cooperation programs.


\begin{thebibliography}{99} 

\bibitem{DLS63} R.H.~Dalitz, R.~Levi Setti, Nuovo Cimento 30 (1963) 489; 
see also L.~Majling, Nucl. Phys. A 585 (1995) 211c. 

\bibitem{davis05} D.H.~Davis, Nucl. Phys. A 754 (2005) 3c, and references 
listed therein, particularly D.H.~Davis, J.~Pniewski, Contemp. Phys. 27 
(1986) 91. 

\bibitem{fingal12prl} M.~Agnello, et al. (FINUDA Collaboration and A. Gal), 
Phys. Rev. Lett. 108 (2012) 042501. 

\bibitem{fingal12npa} M.~Agnello, et al. (FINUDA Collaboration and A. Gal), 
Nucl. Phys. A 881 (2012) 269. 

\bibitem{akaishi99} Y.~Akaishi, T.~Yamazaki, in {\it Physics and Detectors 
for DA$\Phi$NE}, Eds. S.~Bianco et al., Frascati Physics Series Vol. XVI 
(INFN, Frascati, 1999) pp. 59-74. 

\bibitem{akaishi00} Y.~Akaishi, T.~Harada, S.~Shinmura, K.S.~Myint, 
Phys. Rev. Lett. 84 (2000) 3539. 

\bibitem{nemura02} H.~Nemura, Y.~Akaishi, Y.~Suzuki, Phys. Rev. Lett. 89 
(2002) 142504. 

\bibitem{shinmura02} S.~Shinmura, K.S.~Myint, T.~Harada, Y.~Akaishi, 
J.~Phys. G 28 (2002) L1, and a revised calculation by S.~Shinmura, et al., 
Mod. Phys. Lett. A 18 (2003) 128. 

\bibitem{E10} T.~Takahashi, Nucl. Phys. A (HYP2012 Proceedings, in press) 
http://dx.doi.org/10.1016/jnuclphysa.2012.12.118, and J-PARC E10, 
http://j-parc.jp/researcher/Hadron/en/Proposal$\_$e.html. 


\bibitem{millener12} D.J.~Millener, Nucl. Phys. A 881 (2012) 298, 
and references therein.  

\bibitem{tamura10} H.~Tamura, et al., Nucl. Phys. A 835 (2010) 3, 
and references therein. 

\bibitem{GSD71} A.~Gal, J.M.~Soper, R.H.~Dalitz, Ann. Phys. (N.Y.) 63 
(1971) 53. 

\bibitem{millener10} D.J.~Millener, Nucl. Phys. A 835 (2010) 11. 

\bibitem{NSC97} Th.A.~Rijken, V.J.G.~Stoks, Y.~Yamamoto, Phys. Rev. C 59 
(1999) 21. 

\bibitem{umeya11} A.~Umeya, T.~Harada, Phys. Rev. C 83 (2011) 034310. 

\bibitem{GSD78} A.~Gal, J.M.~Soper, R.H.~Dalitz, Ann. Phys. (N.Y.) 113 
(1978) 79. 

\bibitem{MGDD85} D.J.~Millener, A.~Gal, C.B.~Dover, R.H.~Dalitz, 
Phys. Rev. C 31 (1985) 499. 

\bibitem{tamura12} H.~Tamura, et al., Nucl. Phys. A 881 (2012) 310, 
and references therein. 

\bibitem{myint02} K.S.~Myint, Y.~Akaishi, Prog. Theor. Phys. Suppl. 146 
(2002) 599; Y.~Akaishi, K.S.~Myint, AIP Conf. Proc. 1011 (2008) 277; 
Y.~Akaishi, Prog. Theor. Phys. Suppl. 186 (2010) 378; Theingi, K.S.~Myint, 
Y.~Akaishi, Genshikaku Kenkyu 57, Suppl. 3 (2013) 70.  

\bibitem{korshen01} A.A.~Korsheninnikov, et al., Phys. Rev. Lett. 87 (2001) 
092501. 

\bibitem{nue13} S.N.~Nakamura, et al. (JLab E01-011), Phys. Rev. Lett. 110 
(2013) 012502. 

\bibitem{hiyama09} E.~Hiyama, Y.~Yamamoto, T.~Motoba, M.~Kamimura, 
Phys. Rev. C 80 (2009) 054321. 

\bibitem{hiyama13} E.~Hiyama, S.~Ohnishi, M.~Kamimura, Y.~Yamamoto, 
Nucl. Phys. A 908 (2013) 29. 




\bibitem{umeya09} A.~Umeya, T.~Harada, Phys. Rev. C 79 (2009) 024315. 

\bibitem{Hal} D. Halderson, private communication. 

\bibitem{haidenbauer13} J.~Haidenbauer, S.~Petschauer, N.~Kaiser,
U.-G.~Mei{\ss}ner, A.~Nogga, W.~Weise, Nucl. Phys. A 915 (2013) 24.

\bibitem{MDG88} D.J.~Millener, C.B.~Dover, A.~Gal, Phys. Rev. C 38 (1988) 
2700. 

\bibitem{MJ94} J.~Mare\v{s}, B.K.~Jennings, Phys. Rev. C 49 (1994) 2472. 

\bibitem{finelli09} P.~Finelli, N.~Kaiser, D.~Vretenar, W.~Weise, 
 Nucl. Phys. A 831 (2009) 163. 

\end{thebibliography}
\end{document}